\documentclass{article}

\usepackage{arxiv}

\usepackage[utf8]{inputenc} 
\usepackage[T1]{fontenc}    
\usepackage{hyperref}       
\usepackage{url}            
\usepackage{booktabs}       
\usepackage{amsfonts}       
\usepackage{nicefrac}       
\usepackage{microtype}      
\usepackage{lipsum}
\usepackage[numbers,round]{natbib}
\usepackage{amssymb,amsmath}
\usepackage{graphicx}
\usepackage{tabulary}

\title{Analytical and Fast Fiber Orientation Distribution Reconstruction in 3D-Polarized Light Imaging}

\author{
  Abib Alimi\\ corresponding author\\
  Athena Project-Team\\ Inria Sophia Antipolis-M\'editerran\'ee\\ Universit\'e C\^ote d'Azur, France\\ \texttt{stariate@ee.mount-sheikh.edu}
   \And
  Samuel Deslauriers-Gauthier\\ 
  Athena Project-Team\\ Inria Sophia Antipolis-M\'editerran\'ee\\ Universit\'e C\^ote d'Azur, France
   \And
  Felix Matuschke\\
  Institute of Neuroscience and Medicine (INM-1)\\ Research Center J\"ulich, Germany
   \And
  Andreas M\"uller\\
  Simulation Lab Neuroscience\\ J\"ulich Supercomputing Centre\\ Institute for Advanced Simulation\\ JARA\\ Research Center J\"ulich, Germany
   \And
  Sascha E. A. Muenzing\\
  Institute of Neuroscience and Medicine (INM-1)\\ Research Center J\"ulich, Germany
   \And
  Markus Axer\\
  Institute of Neuroscience and Medicine (INM-1)\\ Research Center J\"ulich, Germany
   \And
  Rachid Deriche\\
  Athena Project-Team\\ Inria Sophia Antipolis-M\'editerran\'ee\\ Universit\'e C\^ote d'Azur, France
}

\begin{document}
\maketitle

\begin{abstract}
Three dimensional Polarized Light Imaging (3D-PLI) is an optical technique which allows mapping the spatial fiber architecture of fibrous postmortem tissues, at sub-millimeter resolutions. Here, we propose an analytical and fast approach to compute the fiber orientation distribution (FOD) from high-resolution vector data provided by 3D-PLI. 
The FOD is modeled as a sum of $K$ orientations/Diracs on the unit sphere, described on a spherical harmonics basis and analytically computed using the spherical Fourier transform. 
Experiments are performed on rich synthetic data which simulate the geometry of the neuronal fibers and on human brain data.
Results indicate the analytical FOD is computationally efficient and very fast, and has high angular precision and angular resolution.
Furthermore, investigations on the right occipital lobe illustrate that our strategy of FOD computation enables the bridging of spatial scales from microscopic 3D-PLI information to macro- or mesoscopic dimensions of diffusion Magnetic Resonance Imaging (MRI), while being a means to evaluate prospective resolution limits for diffusion MRI to reconstruct region-specific white matter tracts.
These results demonstrate the interest and great potential of our analytical approach.
\end{abstract}


\section{Introduction}
\label{sec:intro}
To improve our understanding of the human brain structure and function as well as its normal development and complex disorders, it is compulsory to develop and combine different imaging techniques at multiple spatial scales. With a millimeter resolution, diffusion Magnetic Resonance Imaging (MRI) \citep{le1986mr,basser1994mr,tuch2002high,jones2010diffusion,hagmann2006understanding,tournier2019diffusion,leemans2019diffusion} is the only {\it in vivo} and {\it non invasive} imaging technique able to assess the structural architecture of human fibrous organs such as the brain and the heart and to infer their respective connectivity \citep{mori2002fiber,descoteaux2008deterministic,jeurissen2019diffusion,henssen2019ex}. However, {\it in vivo} diffusion MRI suffers from a lack of ground truth \citep{maier2017challenge} and a relatively low spatial resolution. Therefore, 3D-Polarized Light Imaging (3D-PLI) \citep{axer2011novel,jouk2000three,larsen2007polarized,reckfort2015,alimi2017solving} has been investigated and, thanks to its high spatial resolution, it is presented as a complementary and potential technique for the validation and guidance of diffusion MRI fiber orientation estimates and tractography.

3D-PLI is an optical imaging technique that utilizes the birefringence property of brain tissue to map its spatial organization at micrometer resolution \mbox{\citep{axer2011novel,jouk2000three}}. 
The term `3D' in this context is first of all related to the within-section 3D fiber orientation estimation, and only secondly in the sense of 3D volume reconstruction of serial sections. The technique, thus,
provides high-resolution 3D microscopic fiber orientation measurements recovered from each unstained histological tissue section.
Complementing diffusion MRI with 3D-PLI measurements, therefore, requires to bridge the spatial scale information from micro to millimeter dimensions.
Thus, the fiber orientation distribution (FOD), which is widely used in diffusion community \citep{tournier2004direct,alexander2005multiple,tournier2007robust,daducci2014quantitative,dell2019modelling}, is a suitable tool to compare diffusion MRI and 3D-PLI fiber orientation estimates in 3D. 
For validation purpose, 3D-PLI has been used to reconstruct brain white matter (WM) fiber pathways and presented as an adequate means to validate diffusion MRI fiber tractography \citep{alimi2019towards}.
Moreover, \mbox{\cite{caspers2015target,zeineh2016direct,henssen2019ex}} used high-resolution 3D-PLI, or more specifically the generated Fiber Orientation Map (FOM) which reflects the determined 3D fiber courses best, in order to identify and to follow the tiny fiber tracts to complement diffusion MRI results from the same region, rather than addressing long-distance tractography in both modalities.
Concerning the brain's fiber architecture, histology-based studies compared FODs reconstructed from diffusion MRI to those estimated either in 2D from stained tissue sections \citep{leergaard2010quantitative,budde2012examining}, or more recently, in 3D via optical coherence tomography  \citep{wang2015structure} and confocal microscopy \citep{khan20153d,schilling2016comparison}. 
However, to our knowledge, 3D-PLI has not yet been used to evaluate 3D fiber orientation distributions obtained from diffusion MRI, and building PLI-based FODs paves the way for it.

On this account, \cite{axer2016estimating} proposed an estimate of the FOD derived from high-resolution 3D-PLI data, called {\it pliODF}. This technique enables the bridging of spatial scales from micro to macroscopic dimensions, for instance, to the millimeter resolution of diffusion MRI. 
pliODF benefits from the important concept of super-voxel (SV) which allows downsampling of the high-resolution data, and then estimating the FOD on the unit sphere by expanding a normalized directional histogram with a series of spherical harmonics (SH) functions. By changing the SV size, the FOD can be computed at different spatial scales \citep{axer2016estimating,alimi2018analytical,alimi2019analyticalPerformance}. \cite{axer2016estimating} refer to a tissue voxel containing a single high-resolution fiber orientation as `native voxel' and the super-voxel regroups $K = n_x \times n_y \times n_z$ native voxels.
The pliODF, though, is derived from an empirically parameterized directional histogram, which makes it a discrete method. It relies on a histogram binning procedure which can introduce angular errors in the recovered FODs \citep{alimi2018analytical}, and since it is represented on a SH basis, it can also limit the order of its SH expansion \citep{alimi:hal-01659253}. In order to overcome these issues,  \cite{alimi:hal-01659253} proposed to regularize the pliODF using the Laplace-Beltrami operator and therefore improved its results.

In this work, we present an analytical procedure to determine the FOD from 3D-PLI data in order to assess the spatial distribution of fiber orientations within a super-voxel. 
Our method is based on defining each fiber orientation in a native voxel as a 2D Dirac delta function on the unit sphere and the FOD as a sum of these Diracs within each super-voxel. 
The FOD is continuously described on a spherical harmonics basis and analytically computed via the spherical Fourier transform and by means of the Diracs. 
We remind that the FOD is the spatial distribution of the fiber orientations within an image voxel as known in diffusion MRI \mbox{\citep{tournier2004direct,dell2019modelling}}. In the context of 3D-PLI, \mbox{\cite{axer2016estimating}} introduced a discrete approximate of the FOD called the `pliODF' and, here, we propose an analytical FOD denoted `aFOD'.

This paper reviews and extends our previous works in which the analytical method was introduced \citep{alimi2018analytical} and its performance evaluated \citep{alimi2019analyticalPerformance}. 
Here, we give 
(1) a full description of our method to analytically compute the aFOD from high-resolution 3D-PLI vector data, then
(2) the used datasets and the different evaluation criteria are presented for a further/extended comparison with pliODF. 
Results demonstrate
(3) the high angular resolution and precision of the aFOD and the dependence of pliODF on its empirical histogram binning to perform similarly, and
(4) that the aFOD is computationally very efficient and very fast.
We finally 
(5) compute aFODs on human brain data 
at different spatial scales while preserving the original high-resolution organization of WM fiber bundles and 
(6) present it as a means for diffusion MRI to assess potential resolution limits to track specific fascicles in the human brain.

\section{Theory}
\label{sec:theory}
The fiber orientation distribution provides a comprehensive statistical description of integrated 3D-PLI vector data \citep{axer2016estimating,alimi2018analytical,alimi2019analyticalPerformance} and positions itself as a suitable tool for multimodal analysis and comparisons \citep{schilling2016comparison}.
As a function defined on the unit sphere $\mathbb{S}^2$, the FOD can be expanded as a linear combination of spherical harmonics.
In this section, we give a brief background information on spherical harmonics before presenting the pliODF estimate by \cite{axer2016estimating} and the analytical approach --aFOD-- we propose.

\subsection{Spherical harmonics}
The spherical harmonics of order {\it l} and phase factor {\it m} are defined as
\begin{equation} \label{eq:ylm}
Y_l^m(\theta,\phi) = N_l^m P_l^m(\cos\theta)e^{jm\phi}
\end{equation}
where $N_l^m$ 
is a normalization coefficient, $P_l^m$ are the associated Legendre polynomials, the angles $\theta \in [0, \pi] $ and $\phi \in [0, 2\pi)$ are the colatitude and azimuth, respectively. The spherical harmonics $\{Y_l^m:-l\leqslant m\leqslant l, l=0,1,... \}$ form an orthonormal basis over $L^2(\mathbb{S}^2)$ and any square integrable function  $f(\theta, \phi) \in L^2(\mathbb{S}^2)$ can be expressed as
\begin{equation} \label{eq:fsh}
f(\theta,\phi) = \sum_{l=0}^{\infty} \sum_{m=-l}^{l}c_{lm} Y_l^m(\theta,\phi)
\end{equation}
where $c_{lm}$ are the SH coefficients of $f$. These coefficients are obtained by computing the spherical Fourier transform \citep{Healy1998} of $f$ defined as
\begin{equation} \label{eq:coef}
 c_{lm} = \int_{\mathbb{S}^2} f({\bf w})\overline{Y_l^m}({\bf w})d\mathbf{w}
\end{equation}
for ${\bf w} \in \mathbb{S}^2$ and the over-bar denotes the conjugation. The unit vector ${\bf w}$ is related to $(\theta, \phi)$ by
\begin{equation}\label{eq:unit_vect}
{\bf w}(\theta, \phi) = [  \sin \theta \cos \phi  \quad
  \sin \theta \sin \phi  \quad  \cos \theta]^T
\end{equation}
where $\cdot^T$ indicates transposition. In this paper, both functions $f(\theta, \phi)$ and $f(\bf w)$ are assumed to be equal and written interchangeably to simplify notations. 

\subsection{pliODF, a discrete solution}
\cite{axer2016estimating} proposed an estimate of the FOD from high-resolution 3D-PLI vector data, which they called {\it pliODF}. For this, the collection of all native voxels, which represents the fiber orientation map (FOM) \citep{axer2011novel,jouk2000three}, is downsampled into cubic compartments or super-voxels (SV) containing $n_x \times n_y \times n_z$ native voxels each. This allows to determine the spatial distribution of 3D fiber orientations in a given SV. Thus, starting from high-resolution FOMs, as illustrated in Fig.~(3) in \citep{axer2016estimating}, their approach is implemented as follows:
\begin{enumerate}
    \item Define super-voxels of the FOMs measured from unstained histological tissue sections.
    \item Create a normalized directional histogram on the unit sphere from discretized distribution of fiber orientation vectors in each super-voxel.
    \item Approximate the directional histogram using spherical harmonics, see Eq.~\eqref{eq:fsh}.
\end{enumerate}
In practice, the SH approximation is truncated at a maximum order or bandlimit $L_{max}$. Moreover, in the original implementation of the pliODF technique, the SH coefficients are determined with a linear least square method by minimizing $||\mathbf{h}-\mathbf{B}\mathbf{c}||^2$, yielding
\begin{equation}\label{eq:ls_c}
\mathbf{c} = (\mathbf{B}^T\mathbf{B})^{-1}\mathbf{B}^T\mathbf{h}
\end{equation}
where, in matrix format, $\mathbf{B}$ is the basis matrix of the truncated SH, $\mathbf{h}$ the vector entries of the directional histogram, and $\mathbf{c}$ the coefficients vector of the expansion up to order $L_{max}$.

This technique benefits not only from the 3D nature of 3D-PLI dataset but also from the super-voxel which allows the estimation of the FODs at different spatial scales. Nevertheless, the reconstructed pliODF is an empirical and discrete approximation of the FOD. Indeed, it is a directional histogram defined by its bin centers and the number of 3D-PLI fiber orientation vectors falling in each bin is dependent on the bin size or corresponding dihedral angle.
Therefore, the $L_{max}$ of the SH expansion of the pliODF can be limited as shown in \citep{alimi:hal-01659253}.

\subsection{Analytical FOD in 3D-PLI}
We demonstrate, in this section, that without any discretization which leads to loss of orientation information, the aFOD can be analytically reconstructed from high-resolution 3D-PLI orientation measurements. We show how we model our aFOD before we obtain the exact SH coefficients that uniquely characterize it.

\subsubsection{aFOD as sum of k Diracs}
In each voxel of the fiber orientation map (visualizing the dominant fiber orientation), we propose to describe the unit fiber orientation vector as a two dimensional Dirac delta function $\delta$ on the sphere. Therefore, in a super-voxel containing $K$ orientations $(\theta_k, \phi_k)$, the fiber orientation distribution function $f$ can be modeled as the sum of $K$ Diracs \citep{deslauriers2016application,deslauriers2013sampling}, that is
\begin{equation} \label{eq:fdir}
f(\theta, \phi) = \frac{1}{K}\sum_{k=1}^{K} \delta(\cos \theta - \cos \theta_k)\delta(\phi - \phi_k)
\end{equation}
Note that $f(\theta, \phi)$ is completely defined by parameters $(\theta_k, \phi_k)$  of the $K$ orientations located in the SV. In the next step, the coefficients of the SH expansion of $f$ are determined. 
\subsubsection{Computation of the SH coefficients} 
Unlike \cite{axer2016estimating} who used a linear least square method to get the coefficients of the SH expansion, we directly recover the coefficients that uniquely define our aFOD $f$ by computing the spherical Fourier transform through Eq.~\eqref{eq:coef} since we know the parameters $(\theta_k, \phi_k)$ of all fiber orientation vectors determined from 3D-PLI analysis \citep{axer2011novel,jouk2000three,alimi2017solving}
, in a given super-voxel. That is,
\begin{equation*}
\begin{split}
 c_{lm} & = \int_{\mathbb{S}^2} f({\bf w})\overline{Y_l^m}({\bf w})d\mathbf{w} \\
 		& = \int_{0}^{2\pi} \int_{0}^{\pi} f(\theta, \phi)\overline{Y_l^m}(\theta, \phi)\sin\theta d\theta d\phi
\end{split}
\end{equation*}
then $f$ is replaced by the collection of Diracs in Eq.~\eqref{eq:fdir} and $Y_{l}^{m}$ by its expression in Eq.~\eqref{eq:ylm} to get
\begin{equation*}
\begin{split}
c_{lm} & = \frac{N_l^m}{K}\sum_{k=1}^{K} \int_{-1}^{1} \delta(\cos \theta - \cos \theta_k) P_l^m(\cos\theta) d\cos \theta \\
	& \times \int_{0}^{2\pi}\delta(\phi - \phi_k)e^{-jm\phi} d\phi
\end{split}
\end{equation*}
and finally applying the sifting property of the Diracs gives
\begin{align}
 c_{lm} & = \frac{1}{K}\sum_{k=1}^{K} N_l^m P_l^m(\cos\theta_k)e^{-jm\phi_k} \notag \\
 		& = \frac{1}{K}\sum_{k=1}^{K} \overline{Y_l^m}(\theta_k, \phi_k) \label{eq:exact_coef}
\end{align}
This solution for the $c_{lm}$ is the analytical and continuous equivalent of the discrete one approximated for pliODF. Eq.~\eqref{eq:exact_coef} yields, with respect to $L_{max}$, the exact solution which requires no directional histogram and, therefore, all the information from the high-resolution fiber orientation vectors are taken into account in the FOD computation. Furthermore, it can be evaluated at any $L_{max}$. 

Using Eq.~\eqref{eq:fsh}, consequently, the analytical FOD $\mathbf{f}$ is simply computed through matrix-vector multiplication:
\begin{equation}
    \mathbf{f} = \mathbf{B}\mathbf{c}
\end{equation}
with \mbox{$\mathbf{B}$} the SH basis matrix and \mbox{$\mathbf{c}$} the vector of the analytically recovered SH coefficients.

Note that although our reconstruction approach uses the SH to represent the aFOD, it is independent of the SH basis used and here the real and symmetric basis defined in \citep{descoteaux2007regularized} is chosen. In this case, the number of terms of the SH basis of order $l$ is $L = (1/2)(l +1)(l +2)$, where only the even orders are considered.
\section{Methods}
\label{sec:methods}
\begin{figure*}[t!]
\centering
\centerline{
\includegraphics[scale=.7]{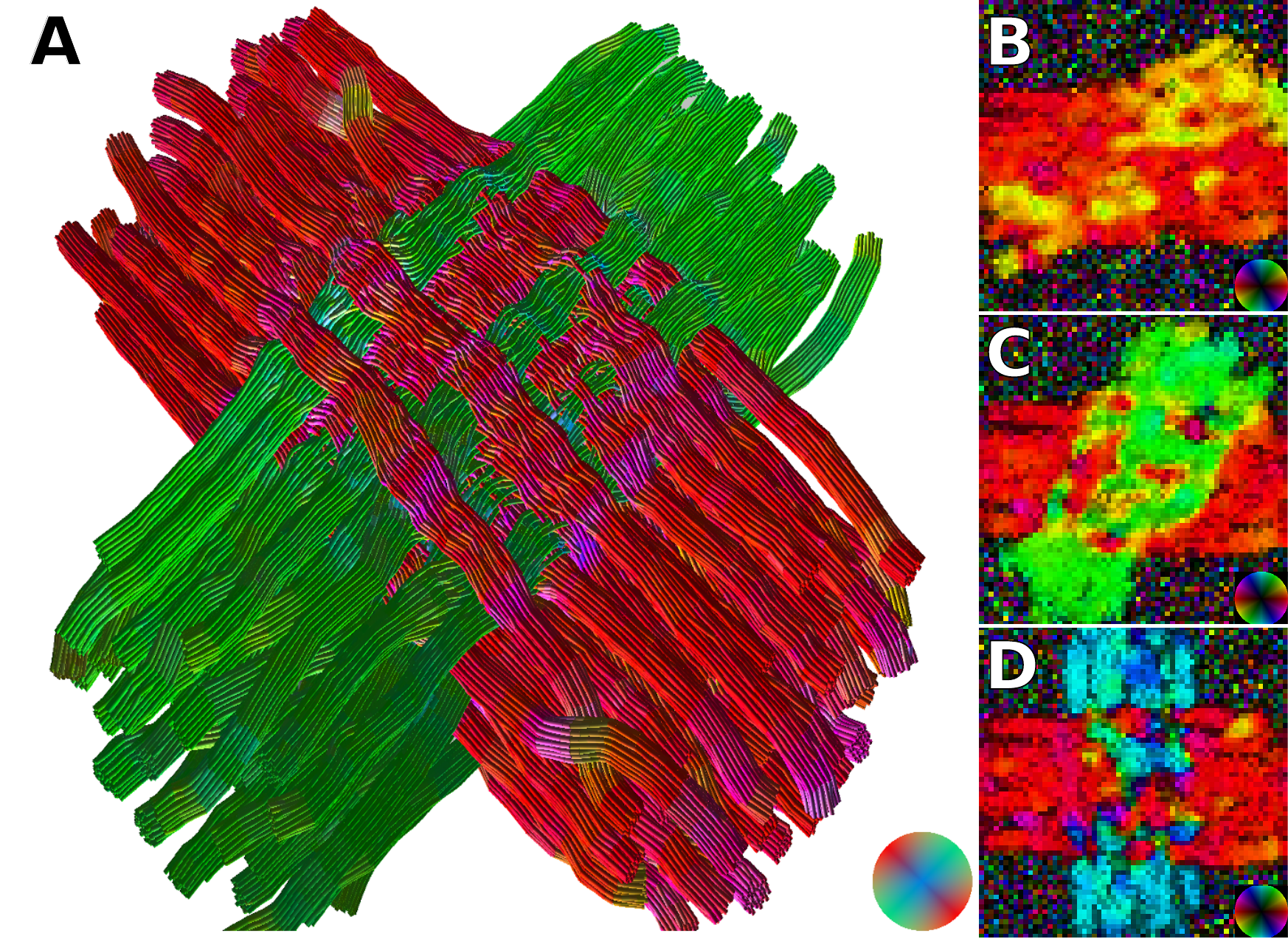}
}
\caption{(A) 3D visualization of the simulated fiber bundles crossing at angle $X~=~90^\circ$ by the FAConstructor \citep{reuter2019faconstructor}. (B)-(D) The resulting Fiber Orientation Maps of 3D-PLI simulations through $60 \text{ }\mu m$ sections with a crossing at $X=30^\circ, 60^\circ$ and $90^\circ$, respectively. The 3D orientations of entire fiber structures (A) and determined local fiber orientations (B)-(D) are color coded as indicated by the color sphere at the bottom right.}
\label{fig:3dcrossing}
\end{figure*}

\subsection{Synthetic data generation}
\label{sec:synth_gen}
The simulated 3D-PLI dataset consists of two 3D fiber bundles intersecting at an angle $X$ in the XY-plane. A 3D collision solving algorithm, described in \citep{matuschke2019}, is performed to ensure that the individual fibers do not overlap. For this purpose, the fibers are divided into short segments so that two segments can be checked for a collision. 
In the event of a collision, the two segments are slowly pushed apart. This happens simultaneously for all colliding objects and is repeated until
there is no more collisions. 
Fig~\ref{fig:3dcrossing}A shows an example of the resulting fiber configuration for the $X=90^\circ$ crossing case. Note that $X$ is varied from $0^\circ$ to $90^\circ$ with $10^\circ$ increments to give $10$ different crossing situations. More details on the fiber generation can be found in \ref{sec:app_congfi2}.

The 3D models generated in this way are then used for the simulation using SimPLI tool \citep{dohmen2015understanding,menzel2015jones} to produce synthetic 3D-PLI datasets. 
The noise of the CCD camera is modelled as in \citep{schmitz2018derivation} i.e. with a negative binomial distribution in which the variance $\sigma$ depends on the expected light intensity $\mu$ by $\sigma=3\mu$.
From the generated 3D-PLI signal, high-resolution orientations are reconstructed based on the approaches in \citep{axer2011novel,schmitz2018derivation}. The generated data consists of 10 slices of $60\text{ }\mu m$ thick, with a pixel resolution of $64\times64 \text{ }\mu m^2.$
\subsection{Experimental brain data}
\label{sec:exp_data}
The post mortem human brain tissue used for this study was acquired in accordance with the local ethic committee of our partner university at the Heinrich Heine University D\"usseldorf. As confirmed by the ethic committee, postmortem human brain studies do not need any additional approval, if a written informed consent of the subject is available. For the research carried out here, this consent is available.

The brain was removed $24 \text{ } h$ after the donor's death (male, age: 54 years, suffered from multiple system atrophy). The right hemisphere was fixed in $4\%$ buffered formaldehyde solution for 4 years. After immersion in a $20\%$ solution of glycerin with Dimethyl sulfoxide (DMSO) for cryoprotection, the brain was deep frozen and cryo-sectioned at $70 \text{ } \mu m$ thickness (with Polycut CM 3500, Leica, Germany) along the coronal plane, orthogonal to the line connecting the anterior and posterior commissures determined from a post mortem turbo spin-echo MRI scan pre-sectioning.
Before each sectioning step, en face images (referred to as blockface images) were taken from the frozen surface of the brain block to generate reference images used for serial section image alignment. Each section was finally mounted unstained on a glass slide, immersed in a $20\%$ solution of glycerin to avoid dehydration, and cover-slipped.

50 consecutive sections were selected for the present study and measured with the Large-area Polarimeter (LAP) setup with tilting stage, as described in \mbox{\citep{axer2011high}}. Pixel size of all images was set to $64 \times 64 \text{ }\mu m^2$. The experimental protocol provided one planar measurement and four tilting position  measurements (in north, west, east, and south direction with $8 \text{ } ^ \circ$ tilting angle), and each measurement was composed of a set of 18 images taken at $10 \text{ } ^ \circ$ equidistant polarization filter rotation angles. Image analysis included calibration, segmentation, and tilting analysis to determine the local fiber structure orientations, utilizing the analysis method described in \mbox{\citep{schmitz2018derivation}}. 

Three-dimensional reconstruction of the brain slab volume consisting of the 50 sections was achieved by a multi-step registration approach: Firstly, the 3D-PLI images were aligned to their corresponding blockface images via affine transformations utilizing in-house developed software based on the software packages ITK \mbox{\citep{yoo2002engineering}} and elastix \mbox{\citep{klein2009elastix}}. Secondly, non-linear registration using the ANTs toolkit \mbox{\citep{avants2011reproducible}} was performed. Finally, all 3D-PLI modalities were spatially transformed using the obtained deformation fields and orientations were corrected accordingly.

One tissue section was measured with the Polarizing Microscope (PM) setup without tilting stage, as described in \mbox{\citep{reckfort2015}}. Pixel size of was set to $1.3 \times 1.3 \text{ }\mu m^2$. The experimental protocol provided just the planar measurement with each measurement being composed of a set of 18 images taken at $10 \text{ } ^ \circ$ equidistant polarization filter rotation angles. Image analysis included calibration, segmentation, and flat image analysis to determine the local fiber structure orientations, utilizing the analysis method described in \mbox{\citep{reckfort2015}}.

\begin{table}
    \caption{Imaging spatial resolutions resulting from super-voxels.}
    \centering
    \begin{tabular}{cc}\toprule
    
     Super-voxel size & Resolution [$mm^3$]\\
    \midrule 
     1 $\times$ 1 $\times$ 1 &  .064 $\times$ .064 $\times$ .07    \\
     2 $\times$ 2 $\times$ 2 &  .128 $\times$ .128 $\times$ .14   \\
     5 $\times$ 5 $\times$ 5 &  .32 $\times$ .32 $\times$ .35   \\
     10 $\times$ 10 $\times$ 10 &  .64 $\times$ .64 $\times$ .7   \\
     25 $\times$ 25 $\times$ 25 &  1.6 $\times$ 1.6 $\times$ 1.75    \\
     50 $\times$ 50 $\times$ 50 &  3.2 $\times$ 3.2 $\times$ 3.5  \\
     \bottomrule
    \end{tabular}
    \label{tab:resolution_full}
\end{table}
\subsection{Evaluations}
To assess the performance of aFOD and pliODF reconstruction approaches and see the relationship between them, we do tests on synthetic and on experimental human brain 3D-PLI data.
\subsubsection{Angular resolution}
\label{sec:ang_resolution}
In diffusion MRI, it is generally assumed that the local maxima or peaks of the FOD coincide with the fiber directions \mbox{\citep{tournier2004direct,tuch2004q,descoteaux2007regularized,daducci2014quantitative}}. Here, the angular resolution corresponds to the critical angle \mbox{$\tau$} under which the {\it peaks} of the two crossing fiber bundles are confounded, that is, they can not be detected separately.
To find the peaks of the FOD, the {\it sh2peaks}\footnote{https://www.mrtrix.org} algorithm implemented in Mrtrix3 \mbox{\citep{tournier2019mrtrix3}} is used,
with the minimum peak amplitude set to 0.5 and the other parameters to default.
\subsubsection{Angular precision}
\label{sec:ang_precision}
Let \mbox{{\it `single fiber'}} denote a bundle of coherently organized/oriented fibers. The angular precision \mbox{$\epsilon$} is the deviation or angular error of the peak of the reconstructed FOD from the simulated ground-truth. 
Here we investigate the \mbox{$\epsilon$} of both aFOD and pliODF, as well as the effect of the directional histogram bin width on \mbox{$\epsilon$}.
The angular precision \mbox{$\epsilon$} is evaluated in degrees as follows,

\begin{equation}
    \epsilon = \frac{180}{\pi M}\sum_{m=1}^{M} \arccos ({\bf v}_m^T {\bf w}_m)
\end{equation}
which is the average distance between \mbox{${\bf v}_m$} the \mbox{$m^{th}$} ground truth orientation and \mbox{${\bf w}_m$} the closest of the recovered peak orientations of the FOD and \mbox{$M$} is the number of fibers in the super-voxel.
\subsubsection{Computational runtimes}
\label{sec:comp_runtimes}
In order to ensure a fair comparison and due to the large size of 3D-PLI human brain data, both reconstruction algorithms are performed on the \textit{JURECA} \citep{jureca} supercomputer at Forschungszentrum J\"ulich, in Germany. The supercomputer consists of 1872 compute nodes and each node is equipped with two Intel Xeon E5-2680 v3 Haswell $24$-core processors  operating at $2.5$ GHz. The considered compute nodes contain $128$ GB of main memory each, and their number is set to $5$, for a total of $120$ cores in order to keep the compute time in an acceptable range. The super-voxel size and the spherical harmonics bandlimit $L_{max}$ are taken as parameters, plus the histogram bin size for the pliODF generation.
\section{Results}
\label{sec:results}
\subsection{Single Fiber Orientation}
\label{sec:results_single}
\begin{figure}[ht]
\centering
\centerline{\includegraphics[scale=.6]{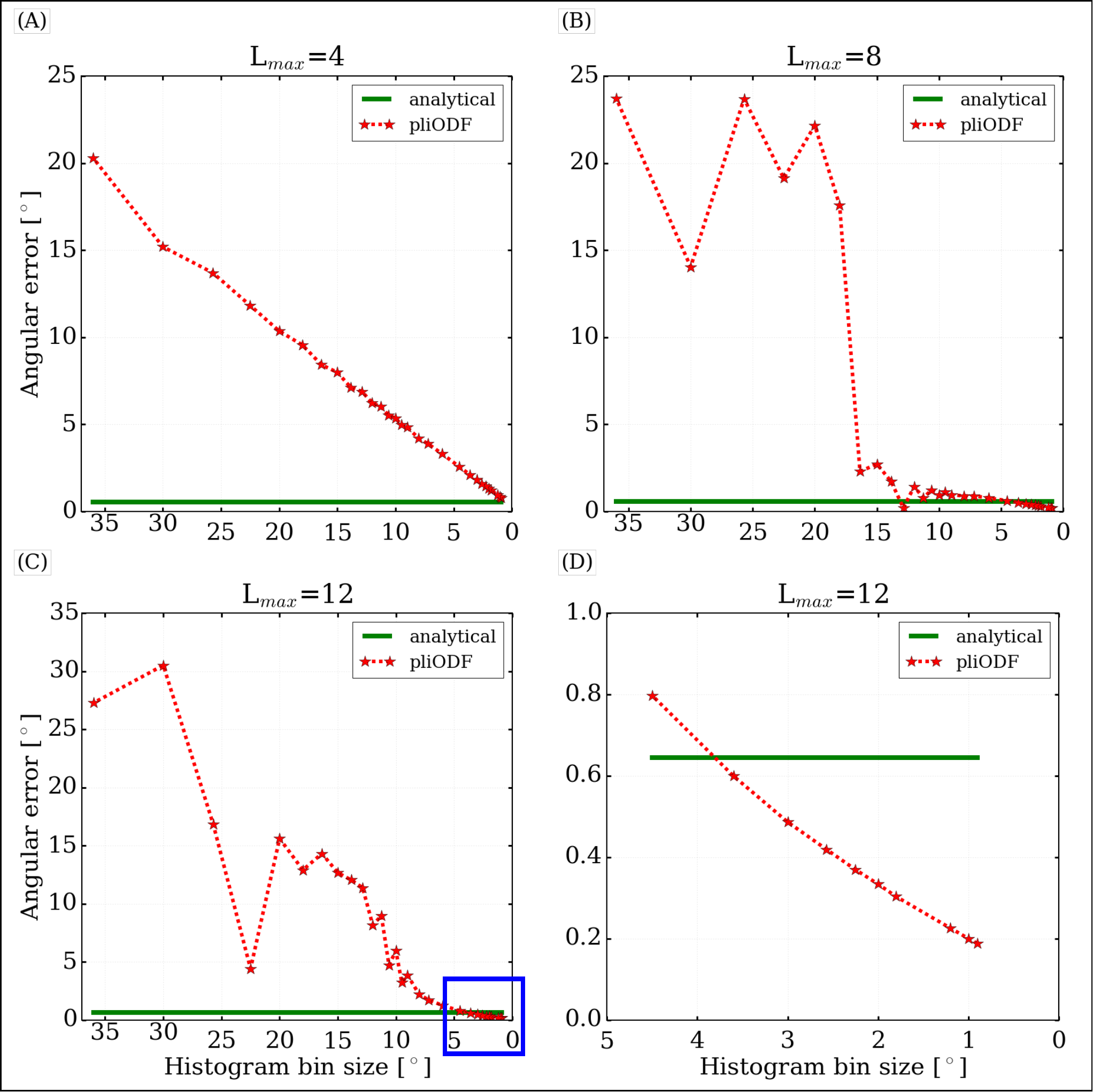}}
\caption{Angular precision $\epsilon$ of `single fiber' FODs as a function of the directional histogram bin size, in degrees. (A)–(C) For decreasing bin size, the $\epsilon$ of pliODF (red dotted line) converges to the constant $\epsilon<1^\circ$ of aFOD, in green horizontal line. At $L_{max}=12$, when bin size is very small, for both FODs $\epsilon<1^\circ$, this is illustrated in (D) where the blue square in (C) is zoomed in.}
\label{fig:single_angular_error}
\end{figure}

We recall that \mbox{{\it single fiber}} refers to a single bundle of fibers. 
Fig.~{\ref{fig:single_angular_error}} depicts, using this simulated single fiber dataset, the FOD angular precision \mbox{$\epsilon$} from both methods as a function of the  directional histogram bin size in degrees. 
The red dotted curve represents the \mbox{$\epsilon$} of pliODF while the green horizontal line is the constant \mbox{$\epsilon$} of aFOD. The \mbox{$\epsilon$} of discrete pliODF converges to the \mbox{$\epsilon$} of aFOD as the bin size decreases, regardless of the considered spherical harmonics \mbox{$L_{max}$}, see (A) to (C). Moreover, in (D) is zoomed in the blue square in (C) to show that with  \mbox{$L_{max}=12$}
and small bin sizes (\mbox{$<5^\circ$}), both pliODF and aFOD have angular precision smaller than \mbox{$1^\circ$} and when the bin size gets smaller, \mbox{$\epsilon$} of pliODF decreases as well. 
Therefore, in the rest of the experiments, in order to get results closer to our analytical approach, bin widths of \mbox{$3.6^\circ$} (consistent with  \mbox{\citep{axer2016estimating}}) and \mbox{$1.8^\circ$} will the considered for the estimation of the pliDOFs.

The previous observations --and choices of bin width-- are corroborated in Table  \mbox{\ref{tab:single_ang_errors}} which quantitatively compares the  \mbox{$\epsilon$} of recovered aFOD and pliODF for the single fiber population with varying SH  \mbox{$L_{max}$}. The angular precision of the aFOD is always  \mbox{$\epsilon <1^\circ$} regardless of  \mbox{$L_{max}$}. This applies to the  \mbox{$\epsilon$} of pliODF, except for  \mbox{$L_{max}=4$} when bin size is  \mbox{$1.8^\circ$}, and  \mbox{$L_{max}\le6$} for a bin width of  \mbox{$3.6^\circ$}. Note, though, that the pliODF produces more erroneous results (e.g. up to  \mbox{$\epsilon > 30^\circ$}) when larger bin width are used (Fig.~{\ref{fig:single_angular_error}}).

\begin{table*}[t]
    \centering
    \begin{tabular}{cp{.5cm}p{1.3cm}cc}\toprule
    
    \multicolumn{5}{r}{Angular Error $\epsilon [^\circ]$}\\ 
    \cmidrule{2-5}
     $L_{max}$& &  aFOD & pliODF $(1.8^\circ)$ & pliODF $(3.6^\circ)$\\
    \midrule 
     4 & & 0.55$^\circ$ & 1.33$^\circ$ & 2.12$^\circ$ \\
     6 & & 0.57$^\circ$ & 0.78$^\circ$ & 1.24$^\circ$ \\ 
     8 & & 0.59$^\circ$ & 0.34$^\circ$ & 0.52$^\circ$ \\
     10 & & 0.62$^\circ$ & 0.03$^\circ$ & 0.09$^\circ$ \\
     12 & & 0.64$^\circ$ & 0.33$^\circ$ & 0.60$^\circ$ \\
     \bottomrule
    \end{tabular}
    \caption{Single fiber angular precision $\epsilon$ in degrees [$^\circ$] as a function of $L_{max}$. 
   For aFODs $\epsilon$ is always $<1^\circ$, and the difference with $\epsilon$ of pliODFs is not very significant with regard to the considered bin sizes.}
    \label{tab:single_ang_errors}
\end{table*}
\subsection{Crossing Fiber Orientations}
\label{sec:results_crossings}
\subsubsection{Angular Resolution}
\label{sec:results_crossing_tau}
\begin{figure}[ht]
\centering
\centerline{\includegraphics[scale=.45]{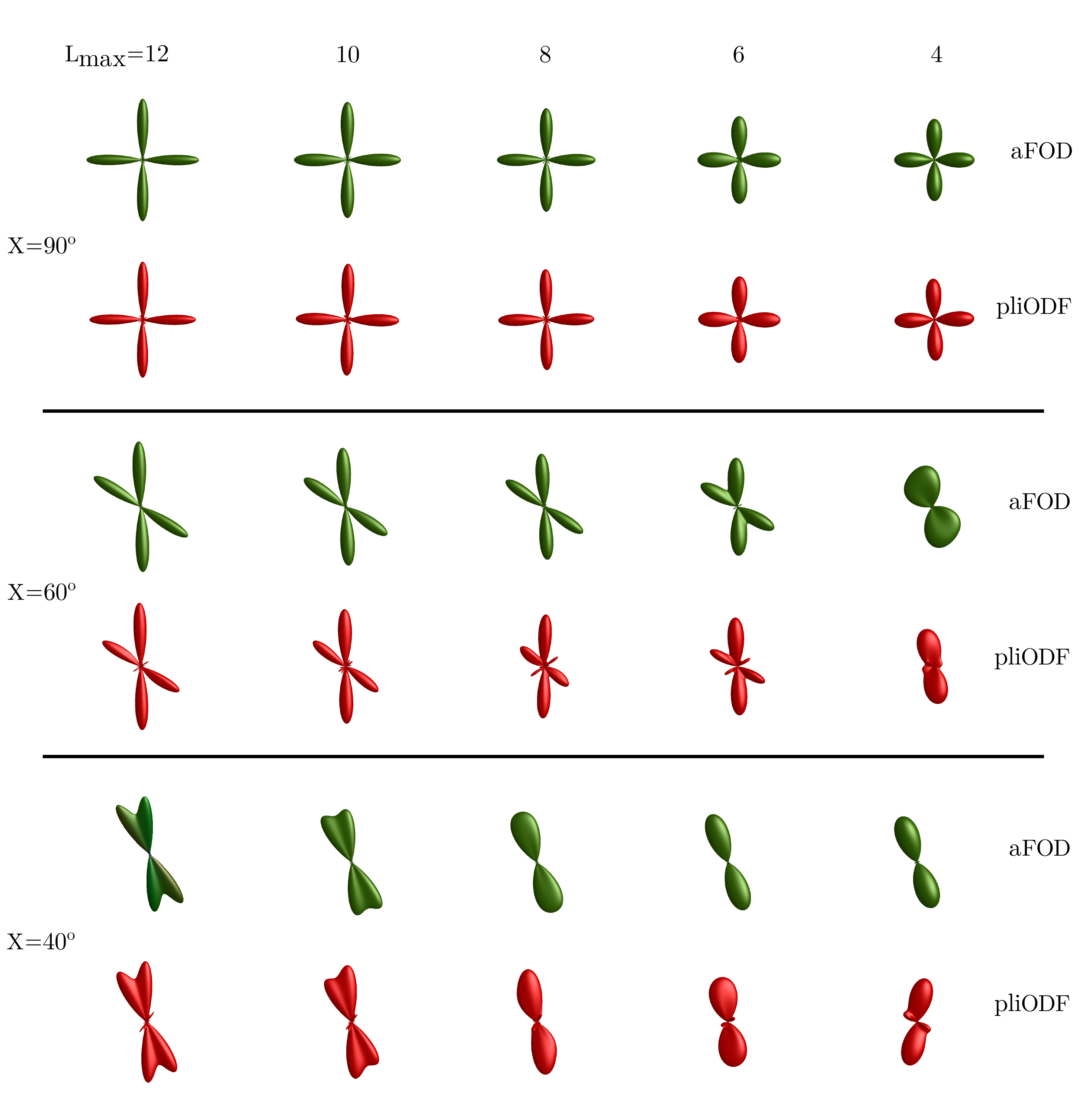}}
\caption{Effects of SH $L_{max}$ on the angular resolution of aFOD and pliODF (with bin size $3.6^\circ$) estimated at different crossing configurations.}
\label{fig:2_fiber_angular_resolution}
\end{figure}

Fig.~{\ref{fig:2_fiber_angular_resolution}} displays the angular resolution of both aFOD and pliODF relative to \mbox{$L_{max}$}. 
It can be observed that \mbox{$90^\circ$}-crossings can be separated at any SH \mbox{$L_{max} \ge 4$}, and at least \mbox{$L_{max}=6$} is necessary to discriminate \mbox{$60^\circ$}-crossings and \mbox{$L_{max}=10$} for \mbox{$40^\circ$}-crossings. 
These observations are valid for the aFOD as well as the pliODF with the selected bin widths, though only results for bin size \mbox{$3.6^\circ$} are shown for illustration. 
Moreover, one can also notice shape differences between the aFOD and the pliODF at \mbox{$40$}- and \mbox{$60^\circ$}-crossings.
\subsubsection{Angular Precision}
\label{sec:results_crossing_epsilon}
\begin{figure}[ht]
\centering
\centerline{\includegraphics[scale=.28]{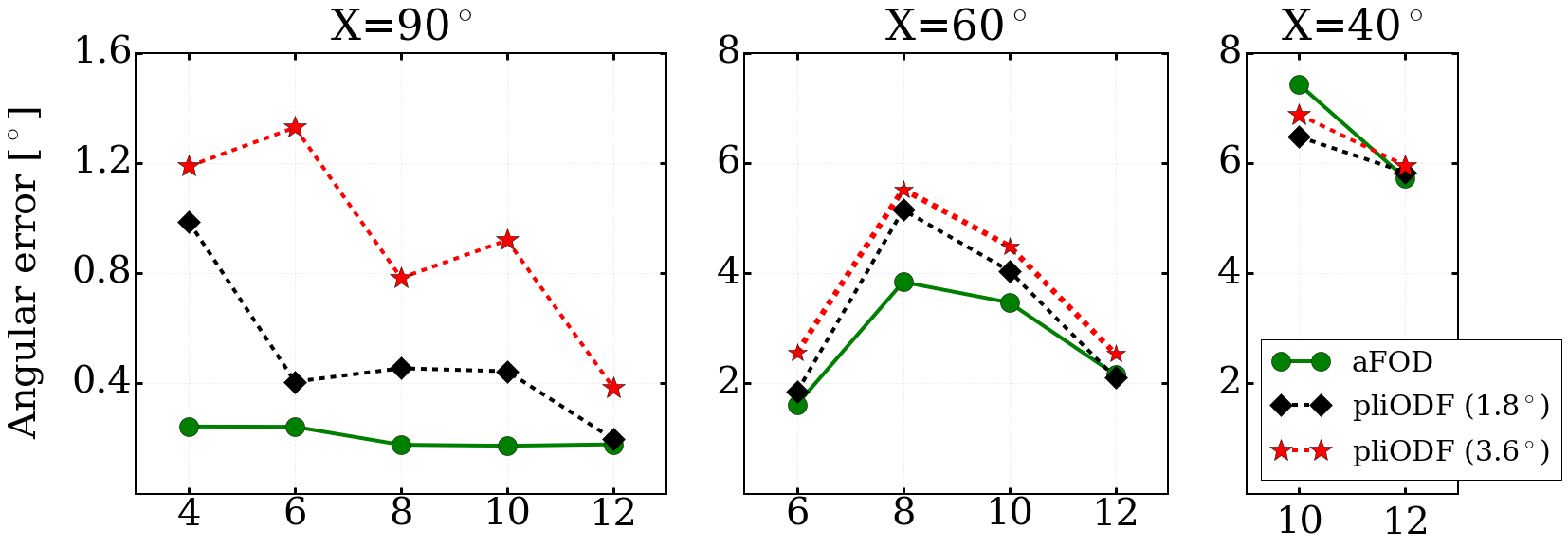}}
\caption{Angular precision of the FODs reconstructed at varying SH $L_{max}$ from different crossing fiber configurations $X$.}
\label{fig:2_fiber_angular_precision}
\end{figure}

Fig.~{\ref{fig:2_fiber_angular_precision}} displays the angular precision \mbox{$\epsilon$} of the FODs reconstructed from crossing fiber bundles using both methods at different spherical harmonics \mbox{$L_{max}$}. 
The green curve represents the \mbox{$\epsilon$} of aFOD and the red and black dotted curves represent the \mbox{$\epsilon$} of pliODF estimated with bin sizes of \mbox{$3.6^\circ$} and \mbox{$1.8^\circ$}, respectively.
The following observations can be made. First, globally for all algorithms, \mbox{$\epsilon$} decreases as the crossing angle $X$ increases: it goes from around \mbox{$7^\circ$} at \mbox{$X=40^\circ$} and \mbox{$L_{max}=10$} down to less than \mbox{$1.5^\circ$} at \mbox{$X=90^\circ$} and \mbox{$L_{max}= 6$}. 
Second, in all combinations of \mbox{$X$} and \mbox{$L_{max}$}, the aFODs have a smaller angular error than pliODFs have, except for crossing \mbox{$X=40^\circ$} and \mbox{$L_{max}=10$} as can be seen in the plot at the right of Fig.~{\ref{fig:2_fiber_angular_precision}}.
Third, as for pliODF, the smaller the bin size, the lower the angular error. Finally, as seen qualitatively in Fig.~{\ref{fig:2_fiber_angular_resolution}} for both methods, the fiber crossing at \mbox{$X=90^\circ$} can be correctly reconstructed at any SH order with a maximum error \mbox{$\epsilon=1.3^\circ$}, the \mbox{$X=60^\circ$} is resolved when \mbox{$L_{max} \ge 6$} with maximum \mbox{$\epsilon <6^\circ$} and, the \mbox{$X=40^\circ$} only when \mbox{$L_{max} \ge 10$} with maximum error \mbox{$\epsilon \approx 7^\circ$}.
\newpage
\subsection{Computational runtime}
\label{sec:results_comp_runtimes}
The FODs' computational runtime measurements on the system \textit{JURECA} \citep{jureca} are presented in Table \ref{tab:running} in minutes:seconds. The test is performed on human brain data consisting of $50$ coronal slices of $1950 \times 1350$ native voxels \footnote{The dataset is actually masked and only $38\%$ of the voxels are considered. These voxels contain the brain tissue.} each (see Sect.~\ref{sec:exp_data}). Concerning pliODF, the directional histogram bin size is set to $1.8^\circ$, equivalent to $20000$ bins in order to get closest results to the analytical solution.
The following observations are made. 
First, the execution time increases with the SH maximum order $L_{max}$ for both reconstruction techniques as previously reported in \citep{axer2016estimating,alimi2019analyticalPerformance}.
Second, at high resolution, the analytical method is faster than the pliODF technique by a factor between $24$ and $73$ at SV size of $5 \times 5 \times 5$ corresponding to $ 320 \times 320 \times 350 \text{ } \mu m^3$ resolution (in Table {\ref{tab:resolution_full}}).
Third, when the SV size is $> 25$ isotropic native voxels, it is $3$ to $6$ times quicker to generate pliODFs than aFODs, and both reconstruction approaches are completed within seconds.
In between these SV sizes, the analytical solution runs faster. 
This could be accounted for the continuous setting of the analytical approach, which makes it compute the SH coefficients from every native voxel located in the super-voxel, regardless of its size. 
However, the discrete pliODF approach estimates the SH coefficients from the defined directional histogram in the considered super-voxel \citep{axer2016estimating,hanel2017interactive}. 
Therefore, pliODF runs moderately faster when the SV size is large, at least $50$ isotropic native voxels since the number of super-voxels in the whole dataset decreases i.e. there is a few directional histograms to estimate the SH coefficients from. Conversely, with a small SV size, the number of super-voxels as well as histograms is large, thus the analytical technique is faster.
\begin{table*}
   \caption{A comparison of runtime measurements in minutes:seconds. 
   All the aFODs are run in the order of seconds regardless of the super-voxel size and the SH $L_{max}$. pliODF takes minutes to run in small sized SV and a few seconds when the resolution is coarser. The execution time generally increases with $L_{max}$.}
    \centering
    \begin{tabular}{lcccccp{1cm}}\toprule
    \multicolumn{6}{r}{Super-voxel size [in voxels]}\\ \cmidrule{3-6}
     & $L_{max}$ & 50 $\times$ 50 $\times$ 50 & 25 $\times$ 25 $\times$ 25 & 10 $\times$ 10 $\times$ 10 & 5 $\times$ 5 $\times$ 5\\
    \midrule 
    pliODF & 4 & 0:01.23 & 0:01.57 & 0:17.29 & 2:07.52 \\
     & 6 & 0:01.90 & 0:02.89 & 0:33.50 & 4:08.71  \\ 
     & 8 & 0:02.92 & 0:04.98 & 0:58.72 & 7:16.63  \\
     & 10 & 0:04.26 & 0:07.71 & 1:30.81 & 11:15.95 \\ 
     & 12 & 0:06.00 & 0:11.10 & 2:12.14 & 16:23.47 \\  
     
    \midrule 
    aFOD & 4 & 0:04.80 & 0:01.27 & 0:01.51 & 0:05.25 \\
     & 6 & 0:09.03 & 0:02.41 & 0:02.34 & 0:07.69 \\ 
     & 8 & 0:15.66 & 0:04.13 & 0:03.73 & 0:09.66 \\
     & 10 & 0:24.97 & 0:06.45 & 0:05.82 & 0:11.38 \\
     & 12 & 0:37.47 & 0:09.70 & 0:07.87 & 0:13.37 \\

     \bottomrule
    \end{tabular}
    \label{tab:running}
\end{table*}
\subsection{On human brain data}
\begin{figure}[h]
\centering
\centerline{\includegraphics[scale=1]{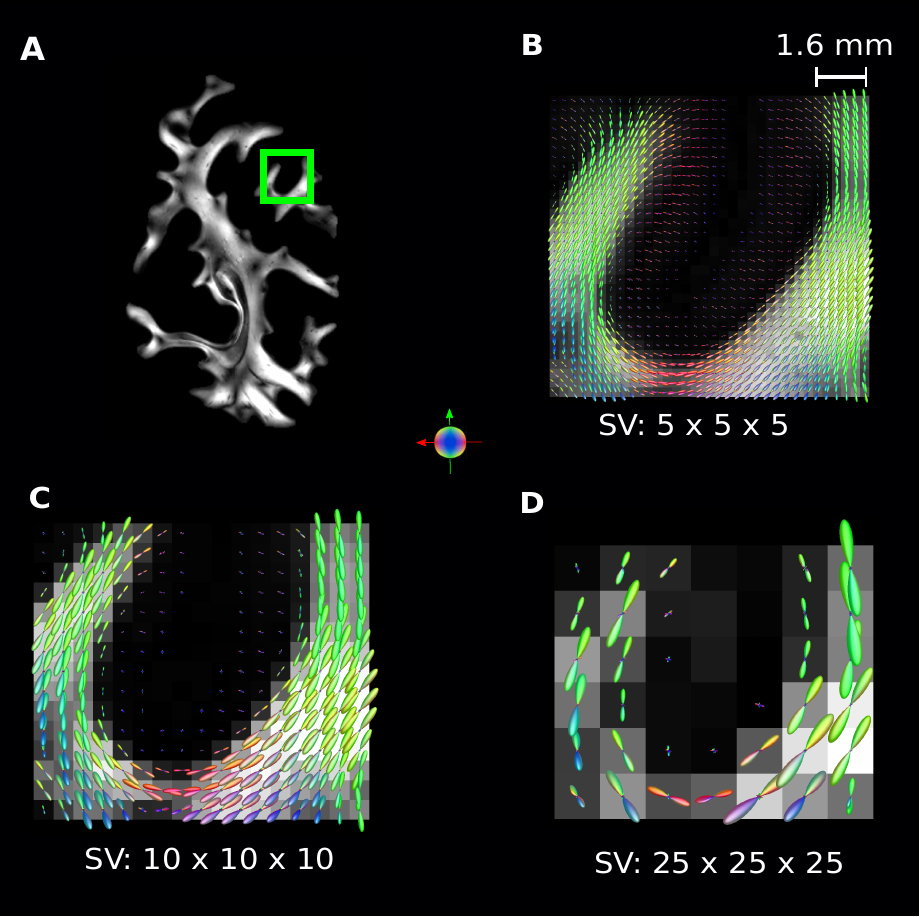}}
\caption{FOD map of U-shaped fibers connecting adjacent gyri at varying resolutions from (B) to (D) with a high-resolution retardation map displayed in (A).  The global fiber organization is preserved through the scales. FODs overlay a recomputed retardation map and are colour-coded according to their local orientation (red: left-right; green: anterior-posterior; blue: inferior-superior).}
\label{fig:u_fibers}
\end{figure}
\begin{figure}[ht]
\centering
\centerline{\includegraphics[scale=.8]{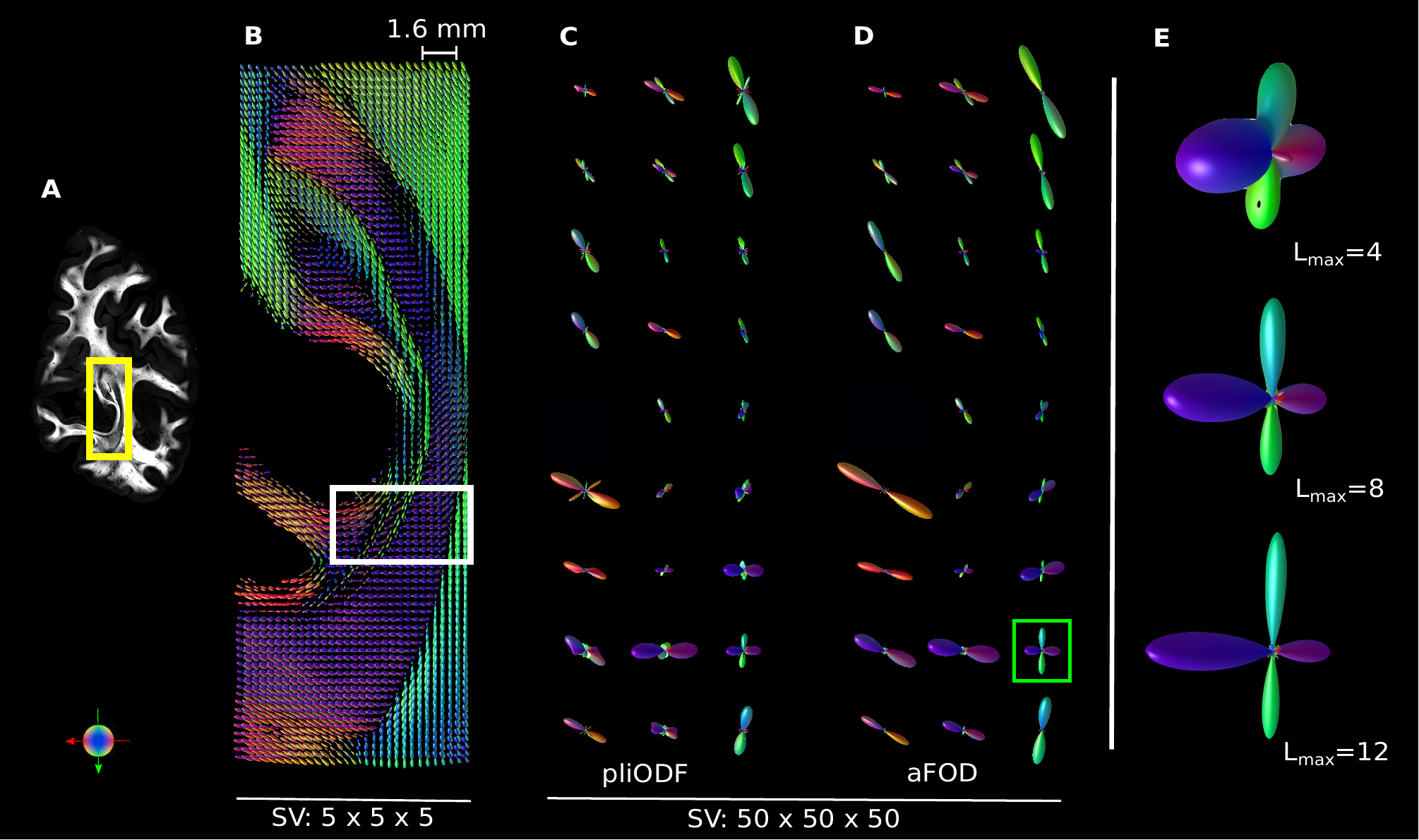}}
\caption{Stratum sagittale (SS) in deep WM described by FODs at different super-voxel sizes i.e. at different spatial resolutions. 
(A) pictures a high-resolution retardation map while (B) to (D) show the SS at SV sizes of $5$ and $50$ isotropic native voxels. 
The global fiber organization is preserved through the scales. FODs are colour-coded according to their local orientation (red: left-right; green: anterior-posterior; blue: inferior-superior). 
The green square RIO in (D) is zoomed in (E) to present a voxel containing a crossing FOD computed at different SH $L_{max}$  and the white rectangular RIO is zoomed in on Fig.~{\ref{fig:res_limit}}.}
\label{fig:stratum}
\end{figure}
The coronal brain slices with a pixel size of $64 \text{ }\mu m$ isotropic presented in Sect.~\ref{sec:exp_data} are used to perform the previous computational running time test, and here,
to illustrate the ability of the analytical method to reconstruct the FODs from real data at different imaging scales while bridging the resolution gap. The corresponding FOM contains $ 1350\times 1950 \times 50$ native voxels \footnote{Again, the dataset is masked.}. 
\subsubsection{U-shaped fibers}
aFODs of U-shaped fibers also known as short arcuate or integrals (in the occipital lobe) are pictured in Fig.~{\ref{fig:u_fibers}}. 
The subplot (A) displays the retardation map \citep{axer2011novel} of a high-resolution section and the green square is the region of interest (RIO).
(B) to (D) overlaying recomputed retardation maps, show the U-shaped fiber connection of neighbouring gyri at different spatial resolutions, from \mbox{$SV=5\times5\times5$} to \mbox{$SV=25\times25\times25$} (see Table~{\ref{tab:resolution_full}}), while preserving their original high-resolution structure.
From these same maps, one can also appreciate the fibers fanning into the gyri.
\subsubsection{Stratum sagittale}
In Fig.~{\ref{fig:stratum}} is shown a FOD map of the stratum sagittale (SS), yellow RIO in (A). The SS consists of longitudinal fibers (going out of the plane), lateral to the ventricles and located in the deep WM region of the occipital lobe.
It can be reported that:
1) Going from smaller \mbox{$SV=5\times5\times5$} (B) to larger \mbox{$SV=50\times50\times50$} (C,D), the spatial resolution gets (\mbox{$10$} times) coarser (\mbox{$3.2\times3.2\times3.5 \text{ } mm^3$}, see Table {\ref{tab:resolution_full}}), however, the global WM tissue structure of the SS is conserved by both the pliODF (C) and the aFOD (D) with crossings.
2) Overall, the shapes of the aFOD and pliODF are very similar, although the latter sometimes presents some spurious lobes at the center of the crossings.
3) The aFOD is more sharply defined as the spherical harmonics \mbox{$L_{max}$} increases as shown in (E). This observation is also valid for pliODF (but not displayed, with eventual small spurious lobes).
A zoom in on the white RIO is displayed in Fig.~{\ref{fig:res_limit}}.
\subsubsection{Diffusion MRI resolution limit}
\begin{figure}[t!]
\centering
\centerline{\includegraphics[scale=.8]{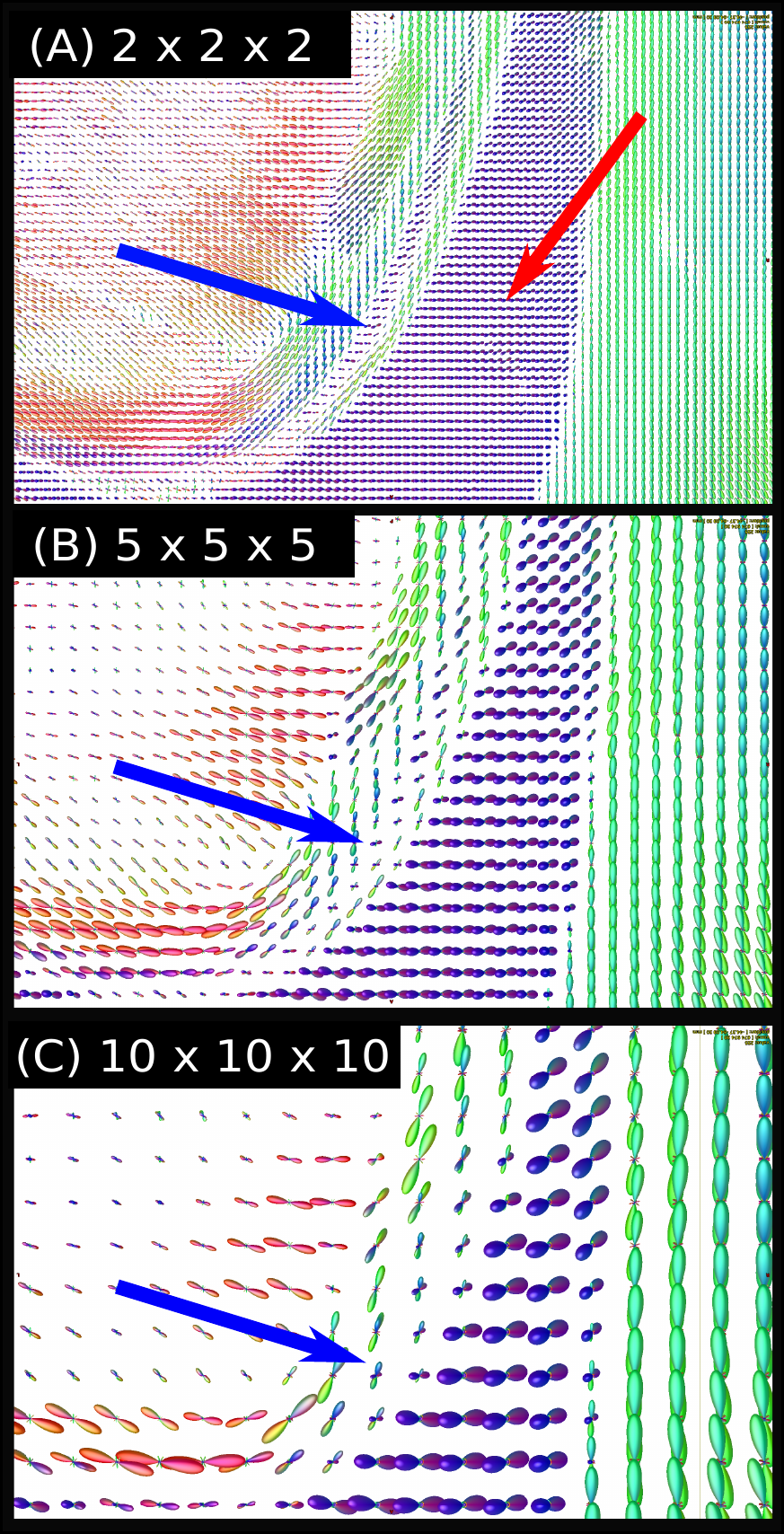}}
\caption{Zoom in on the white rectangle of Fig.~\ref{fig:stratum}: details about the layers of the stratum sagittale. The spatial resolution is $ 128 \times 128 \times 140 \text{ } \mu m^3$ in (A), $ 320 \times 320 \times 350 \text{ } \mu m^3$ in (B) and $ 640 \times 640 \times 700 \text{  } \mu m^3$ in (C). The blue and the red arrows point the thinner and the more pronounced layers of the SS, respectively. By reducing the scale, the thinner layer form crossing aFODs with a vertical fiber bundle.
}
\label{fig:res_limit}
\end{figure}
Fig.~{\ref{fig:res_limit}} displays a zoom in on the aFOD map of the stratum sagittale. At high resolution (A), the two layers of the SS are distinguishable as shown  by \cite{vergani2014intralobar} in their postmortem dissection of the occipital lobe, even though they mentioned the difficulty to appreciate them on coronal sections. From our FOD maps build from high-resolution 3D-PLI vector data, one can separately identify them up to a coarser resolution of $640 \times 640 \times 700 \text{ } \mu m^3$ in (C) where the thinner layer intersect another vertical fiber bundle to form crossing FODs. 
We can see loss of WM fibers at this scale, suggesting, thus, that higher resolutions of $128 \times 128 \times 140 \text{ } \mu m^3$ and $320 \times 320 \times 350 \text{ } \mu m^3$ 
could be considered as suitable spatial scales for diffusion MRI to reach and apply bundle-specific tractography \citep{rheault2019bundle} of the SS. Indeed, using a SH $L_{max} = 6$ at these particular scales, the specific WM tracts of the SS can be tracked, maybe more preferably at SV size of $2 \times 2 \times 2$ equivalent to $128 \times 128 \times 140 \text{ } \mu m^3$ resolution as observed in our study. 
\begin{figure}[t!]
\centering
\centerline{\includegraphics[scale=.24]{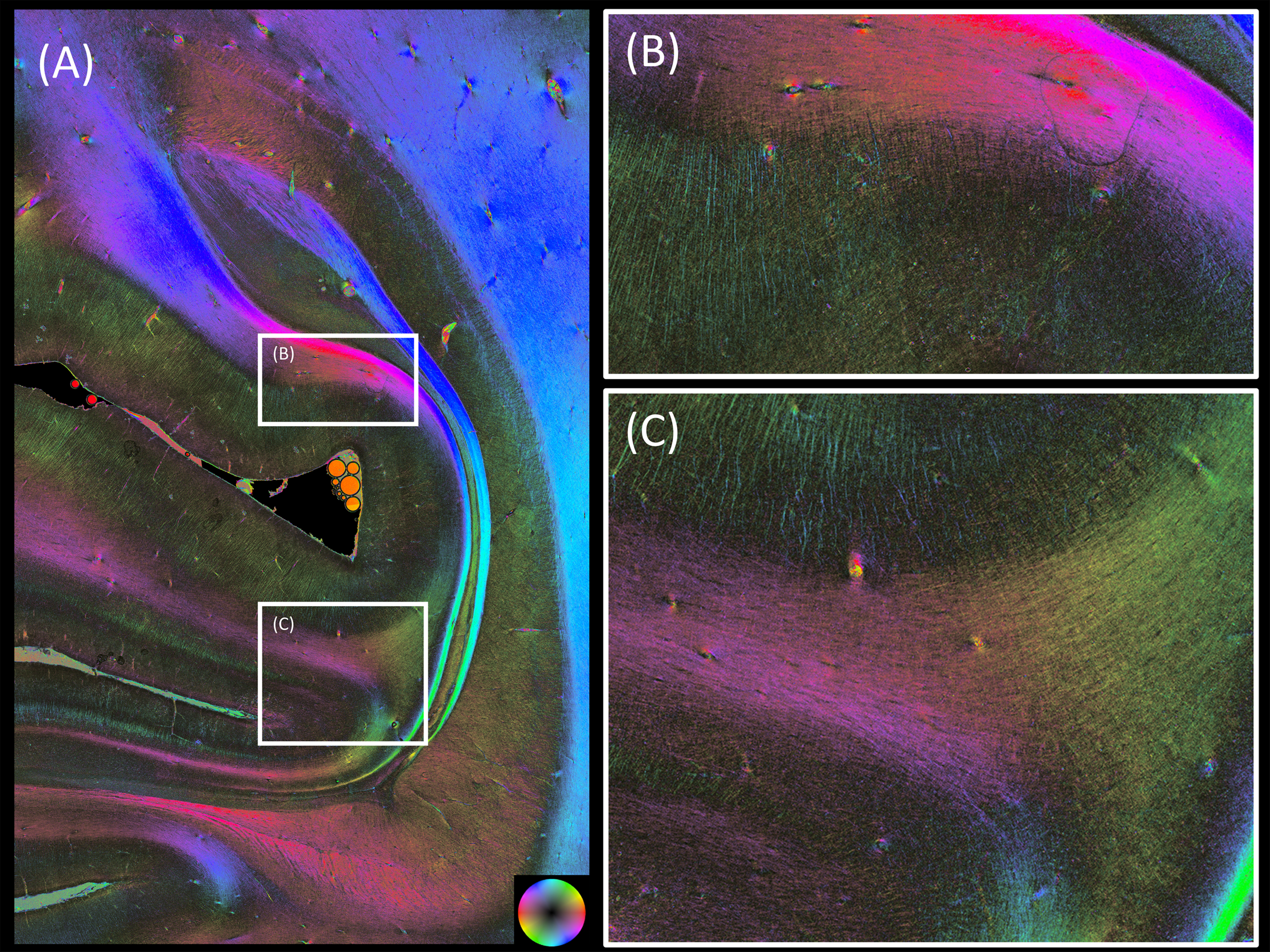}}
\caption{Fiber orientation map of a coronal brain section in the region of the stratum sagittale (A) observed with 3D-PLI at $1.3 \times 1.3 \times \mu m^2$ resolution. Details of crossings (radial and horizontal fibers) in a transition zone between white and gray matter is shown in (B). (C) highlights the crossing of three pathway crossings at the bottom right (dark spot).}
\label{fig:high_res_pli}
\end{figure}
\section{Discussion}
\label{sec:discussion}
In this paper, we propose an analytical and fast approach to resolve the FOD reconstruction problem in 3D-Polarized Light Imaging. 
We quantitatively compare the performances of our approach and the discrete pliODF method on both highly rich synthetic and human brain 3D-PLI datasets.

\paragraph{{\bf On Angular Precision, Angular Resolution}}
As pointed out by \mbox{\citep{parker2013pitfall}}, ``characterisation of any error in the resolution of the single fiber orientation will provide an easy way of identifying any underlying systematic problems inherent in certain (FOD reconstruction) techniques...''. 
Thus, both reconstruction techniques are performed on a \mbox{{\it single fiber}} synthetic model. 
Results show that, unlike our aFOD, pliODF is dependent on the number bins of its discretized directional histogram. This demonstrates that pliODF requires high number of bins to reach results similar to the analytical solution. Indeed, in this experiment (see Sect.~{\ref{sec:results_single}}), the angular precision \mbox{$\epsilon$} of aFOD is always smaller than \mbox{$1^\circ$} while the \mbox{$\epsilon$} of pliODF converges to the aFOD's by decreasing its bin width.

Our aFOD is analytically defined, therefore, the produced angular errors in the single and crossing fiber experiments could be explained by the truncation of the spherical harmonics basis as reported in the previous chapter (and by potential biases from the FOD peaks detection method). In the multiple fiber context, except for \mbox{$X=60^\circ$}-crossing where \mbox{$\epsilon$} increases when \mbox{$L_{max}$} goes from 6 to 8, the angular precision improves when the SH truncation order increases, independent of the crossing configuration. 
In the particular cases of crossing at \mbox{$X=0^\circ \text{ and } X=90^\circ$}, the deviation is always very low, i.e. less than \mbox{$1^\circ$}. 
Moreover, using high SH orders helps sharpen the aFOD (and pliODF with proper bin width) and gives more defined peaks very useful in fiber-tracking applications \mbox{\citep{tournier2019diffusion}}.

As for pliODF, the presence of angular error can be attributed to the SH basis truncation as well, however, it could be mainly due to its discrete nature. 
pliODF by definition uses an arbitrary discretized directional histogram and only provides an approximate of the FOD \mbox{\citep{axer2016estimating}}, which is the main limitation of the approach.
Indeed, the directional histogram is defined by its bin centers and the number of native orientation vectors falling in each bin is dependent on the bin size or corresponding dihedral angle.  
This could explain the fact that, in the single fiber experiment when the histogram bin size is \mbox{$<4^\circ$}, pliODF has lower \mbox{$\epsilon$} than aFOD does. In fact, the closer this bin center to the orientation of single fiber ground truth, the smaller the angular error of pliODF. Conversely, the further the bin center, the larger the \mbox{$\epsilon$} as shown in Fig.~{\ref{fig:single_angular_error}} and Table~{\ref{tab:single_ang_errors}}. In the crossing fiber experiment, results in Fig.~{\ref{fig:2_fiber_angular_precision}} show again that the aFOD presents less angular deviation than the pliODF.

The use of the discretized directional histogram has also a significant influence on the angular resolution of the pliODF in the crossing fiber experiment, 
but the bin sizes we have selected (\mbox{$3.6^\circ \text{ and }1.8^\circ$}) are `empirically' chosen so as to get pliODF results close to the aFOD solution. As said earlier, when the bin width gets smaller, i.e. the number of histogram bins gets larger, pliODF is expected to behave as well as our the aFOD (Fig.~{\ref{fig:2_fiber_angular_resolution}}, Fig.~{\ref{fig:2_fiber_angular_precision}}). 
Nevertheless, one should be cautious about possible over-binning of the directional histogram which could eventually cause some undesirable effects such as the appearance of spurious lobes (Fig.~{\ref{fig:2_fiber_angular_resolution}}, Fig.~{\ref{fig:stratum}}) or eventual extended computation runtimes.

As aforementioned, the value of the spherical harmonics order \mbox{$L_{max}$} determines the maximum achievable FOD angular resolution. By increasing \mbox{$L_{max}$}, more acute crossing fiber populations could be resolved as shown in Sect.~{\ref{sec:results_crossings}}. Here again we set the directional histogram bin size and both approaches give very similar angular resolutions \mbox{$\tau$} but aFOD in general demonstrates better angular precision \mbox{$\epsilon$} (Fig.~{\ref{fig:2_fiber_angular_precision}}).

As for imaging noise influence, in a previous study \mbox{\citep{alimi2019analyticalPerformance}}, we tested our analytical method on a synthetic dataset and it was revealed very little influence of noise on both the angular resolution and angular precision of the aFOD. We expect pliODF to perform as well, if the good number of radial histogram bins is used. 
Indeed, the noise in 3D-PLI data we deal with comes mainly from the CCD camera and its removal is performed during the estimation of the high-resolution fiber orientation vector \mbox{\citep{axer2011novel,axer2011high,kleiner2012classification,wiese2014polarized,alimi2017solving,schmitz2018derivation}}.

\paragraph{{\bf On computational runtimes}}
The analytical approach is very fast and computationally efficient. This could be due to the choice of defining each fiber orientation as a $2D$ Dirac function on the unit sphere, which simplify and accelerate the computation of the SH coefficients, analytically. Note that in our study, only a subset of the human brain was analysed.
An entire human brain of about $1200 \text{ } cm^3$ has, however, about $100$ times more voxels when scanned at $64 \times 64 \times 70 \text{ } \mu m^3$ resolution and it would theoretically last about $100$ times longer to calculate the FODs. Moreover, at a (ultra) higher resolution of $1.3 \times 1.3 \times 70 \text{ } \mu m^3$ the dataset gets about $2 \times 10^5$ times larger and the computations will require more time.
We recall that to reach millimeter resolutions, the SV size should be very large, therefore, pliODF runs faster as seen in Sect.~{\ref{sec:results_comp_runtimes}} and might still be applicable, given a suitable bin size is used. However, since the calculation time at this scale is already quite fast (seconds for the current dataset), the use of the aFOD is still advocated, since it gives the exact solution and has no bias of the histogram binning.

\paragraph{{\bf Limitation}}
One limitation of representing the fiber orientation distribution in a truncated SH basis is the presence of Gibbs ringing effects as discussed in a recent review by \cite{dell2019modelling}. However, this issue can be resolved or drastically mitigated for the aFOD by introducing an apodized delta function in its definition as recently proposed by \cite{raffelt2012reorientation} and \cite{dhollander2014track} in diffusion MRI.

\paragraph{{\bf On human brain data}}
Finally, our analytical algorithm preserves the global integrity of the WM fiber architecture in the human brain. Indeed, experiments in deep WM region (Fig.~{\ref{fig:stratum}}) as well as in the cortex (Fig.~{\ref{fig:u_fibers}}) where the fiber structures are well known \citep{sachs1892hemispharenmark,vergani2014intralobar}, show the conservation of the general anatomical organization of the stratum sagittale (SS) and the U-fibers at different spatial resolutions.  
It is noticed though that pliODF sometimes presents some spurious lobes, which could be due to the empirical (over- or under-) binning of the directional histogram.
Conversely, high SH orders improve the sharpness of the FOD as well as its angular resolution and precision as demonstrated in the simulations (Fig.~{\ref{fig:2_fiber_angular_resolution}}, Fig.~{\ref{fig:2_fiber_angular_precision}}), and commonly known in diffusion MRI \mbox{\citep{tournier2007robust,descoteaux2008high}}.

Further analysis of the SS illustrates the loss of its thinner layer or simply its intersection with a differently oriented fiber population to form more complex fiber configurations. This is visible at a coarse resolution of about $640 \times 640 \times 700 \text{  } \mu m^3$ in Fig.~{\ref{fig:res_limit}}C. Whereas at higher resolutions of around $128 \times 128 \times 140 \text{  } \mu m^3$, both layers can be clearly appreciated. This observation would suggest a potential spatial resolution limit, for diffusion MRI to reach, at which the WM fibers of the SS could be reconstructed and possibly improve their tracking.

Note, that the state-of-the-art 3D-PLI signal analysis only provides a single orientation estimate, i.e. one orientation vector per native voxel. Consequently, an initial image resolution of $64 \times 64 \text{ }\mu m^2$ leads to fiber orientation descriptions (including their fiber tract crossings) at larger spatial scale. This does not mean, that there is no crossing present in human brain data shown in Figs.~({\ref{fig:u_fibers}},{\ref{fig:stratum}},{\ref{fig:res_limit}}). As demonstrated in Fig.~{\ref{fig:high_res_pli}}, the studied region of interest in the sagittal stratum, for example, is full of fiber crossings which are revealed by 3D-PLI at axonal resolution, i.e. at $1.3 \times 1.3 \text{ }\mu m^2$ pixel size. Future research will certainly address aFOD determination in serial 3D-PLI measurements at the one micrometer scale.

\section{Conclusion}
\label{sec:outro}
We have presented an analytical and fast method to define and reconstruct the fiber orientation distribution in 3D-PLI.
To achieve it, we model each high-resolution fiber orientation vector contained in the super-voxel as 2D Diracs delta function on the unit sphere and the aFOD as their sum. 
The aFOD is analytically described on a spherical harmonics basis and efficiently computed via the spherical Fourier transform and by means of the Diracs. This makes the aFOD not only computationally very fast but also with high angular resolution and precision and the ability to close the resolution gap with diffusion MRI.
This work serves as a important step towards a major goal which is the validation of diffusion MRI fiber orientation estimates and histology guided tractography via 3D-Polarized Light Imaging technique, more specifically, it leads to multimodal brain imaging analysis across the scales, through the fiber orientation distribution.

\section*{Acknowledgements}
This work was partly supported by ANR "MOSIFAH" under ANR-13-MONU-0009-01, the ERC under the European Union's Horizon 2020 Research and Innovation Program (ERC Advanced Grant agreement No.\ 694665:CoBCoM).
M.S, F.M and S.M were partly funded by the European Union's Horizon 2020 Research and Innovation Program under Grant Agreement No.\ 785907 (HBP SGA2). 
The authors gratefully acknowledge the computing time granted through JARA-HPC on the supercomputer \textit{JURECA} \citep{jureca} at Forschungszentrum  J\"ulich.

\appendix
\section{Synthetic Data Generation}
\label{sec:app_congfi2}
The nerve fiber model imitates two fiber bundles in the XY-plan crossing at an angle $X$ which varies from  $0^\circ$ to $90^\circ$, with $10^\circ$ increments. To build this configuration, the following steps were performed:
\begin{enumerate}
    \item Two inital fibers bundles with a length of $3840 \text{ } \mu m$ and a radius of $640 \text{ } \mu m$ were placed at an crossing angle $X$.
    \item These initial fiber bundles were then filled with smaller fiber bundles with a radius of $64 \text{ } \mu m$, on which the collision algorithm \citep{matuschke2019} was performed for $25$ steps, to allow a `macroscopic' weave pattern.
    \item The resulting fiber bundles were then filled one last time with fibers with a radius of $10 \text{ } \mu m$. In this state, each fiber is spatially divided into $20 \text{ } \mu m$ segments. To limit the curvature of each fiber, the minimum radius of a bending fiber is limited to $40 \text{ } \mu m$.
\end{enumerate}
This configuration runs with the collision solving algorithm until no collision are detected.

The resulting fiber configuration, for instance as pictured in Fig.~\ref{fig:3dcrossing}A where  $X=90^\circ$, is then used for the 3D-PLI simulation to generate the resulting FOM. For the simulation, the model is discretized on a volume of $3840 \times 3840 \times 600 \text{ } \mu m^3$ to allow the calculations performed by \textit{simPLI} \citep{dohmen2015understanding, menzel2015jones}. Each fiber is modelled as a filled cylindrical object with an absorption coefficient $\mu= 1/{mm}^3$ and a birefringence $\Delta n=-0.001$. The deriving microscopic images have a in-plane resolution of $60 \times 60 \text{ } \mu m^2$. The signal noise was calculated with a gain factor of $g = 3$ (see \citep{schmitz2018derivation}) and an initial light intensity of $26000 \text{ } a.u$. The signal from the 3D-PLI pipeline is then analyzed to calculate the FOM.

\bibliography{main}

\end{document}